\documentclass[pre,aps,twocolumn]{revtex4}
\usepackage{graphicx}
\def\strutdepth{\dp\strutbox}
\def\nw#1{\strut\vadjust{\kern-\strutdepth\vtop to0pt{\vss\hbox to\hsize
{\hskip\hsize\hskip5pt$\leftarrow$\hss\strut}}}{\em #1}}
\usepackage[usenames]{color}
\begin{document}

\title{Maximum size of drops levitated by an air cushion}

\author{Jacco H. Snoeijer$^{1,2}$, Philippe Brunet$^{3}$ and Jens Eggers$^1$}

\affiliation{
$^{1}$Department of Mathematics - University of Bristol, University Walk,
Bristol BS8 1TW, United Kingdom \\
$^{2}$Physics of Fluids Group and J. M. Burgers Centre for Fluid Dynamics,
University of Twente, P.O. Box 217, 7500 AE Enschede, The Netherlands \\
$^{3}$Laboratoire de M\'ecanique de Lille UMR CNRS 8107, Boulevard
Paul Langevin 59655 Villeneuve d'Ascq Cedex, France \\
}

\date{\today}

\begin{abstract}
Liquid drops can be kept from touching a plane solid surface by
a gas stream entering from underneath, as it is observed for water
drops on a heated plate, kept aloft by a stream of water vapor.
We investigate the limit of small flow rates, for which the size
of the gap between the drop and the substrate becomes very small.
Above a critical drop radius no stationary drops can exist, below
the critical radius two solutions coexist. However, only the solution
with the smaller gap width is stable, the other is unstable.
We compare to experimental data and use boundary integral simulations
to show that unstable drops develop a gas ``chimney'' which breaks
the drop in its middle.
\end{abstract}

\maketitle

\section{Introduction}

Drops levitated on an air cushion have numerous applications,
and have generated interest for a long time. For example, in
lens manufacture drops of molten glass can be prevented from contact
with a solid substrate \cite{Duchemin05}. This is achieved by levitating
the glass above a porous mold, through which an air stream is forced.
A second example is the so-called 'Leidenfrost' drop \cite{Leidenfrost},
a drop of liquid on a plate hot enough to create a film of vapor
between the drop and the plate \cite{Holter52,Goldshtik86,Biance03,Linke06}.
Since the drop is thermally isolated insulated by the vapor film, it
can persist for minutes \cite{Biance03}. Finally, a thin air film
is believed to play a crucial role for the ``non-coalescence'' of
a liquid drop bouncing off another liquid surface
\cite{CouderNature,Gilet07,Yacine01}.

The question we will address in this paper is whether for a given
set of parameters, in particular the radius of the drop as it ``rests''
on the substrate, a stationary solution exists and whether it is
stable. Apart from lense manufacture \cite{Duchemin05}, this
question is important for the manipulation of corrosive substances
\cite{Hervieu01} or the frictionless displacement of drops \cite{Linke06}.
Of particular interest is the maximum drop size that can be
sustained, and the limit of very small flow rates. The drop
continues to levitate in this limit since the gap between the
liquid and the substrate becomes very small, so the lubrication
pressure produced by the viscosity of the gas becomes significant.
This enables us to employ asymptotic methods, making use of the
disparity of scales between the gap size and that of the drop.

Experimentally, it is observed that the stability limit is reached
when the radius equals at least a few capillary lengths
$\ell_c=\sqrt{\gamma/(\rho g)}$. This natural length scale for
our system is determined by the surface tension $\gamma$,
density $\rho$ of the liquid, and acceleration of gravity $g$. At
a few capillary lengths, the drop is flattened to a pancake shape.
Biance et al. \cite{These_Biance,Biance03} observed a critical radius

\begin{equation}\label{eq.exp}
\frac{r_{\rm max}}{\ell_c} = 4.0 \pm 0.2,
\end{equation}
where $r_{\rm max}$ is defined in Fig.~\ref{fig:outer}. Beyond this
radius, ``chimneys'' appeared, i.e. bubbles of air trapped below the
curved and concave surface of the drop that rise owing to buoyancy,
and eventually burst through the center of the drop. This suggests that
the critical radius is related to the Rayleigh-Taylor instability
of a heavy fluid (the drop) layered above a light fluid (the gas layer).
In \cite{These_Biance,Biance03} this idea is used to estimate
$r_{\rm max}/\ell_c \approx 3.83$.

While this is close to the experimental value, the argument ignores
the gas flow responsible for the levitation force. This flow was taken
into account by Duchemin et al. \cite{Duchemin05}, who calculated the static
shape of a drop levitated above a {\it curved} porous mould, using
a combination of numerics and asymptotic arguments. For large
drop volume, they found no physical solutions, while for smaller drops
multiple solutions were calculated numerically. Duchemin et al.
\cite{Duchemin05} did not present a formal stability analysis of these
solutions, but suggested that the limit of stability is related to
the appearance of large oscillations on the underside of the drop.

A large number of studies of Leidenfrost drops have focused on
the appearance of self-sustained oscillations of
the drop \cite{Holter52,Watchers66,Takaki84,Takaki85,Tokugawa94,
Strier00,Biance03,Aranson08}. These oscillations can sometimes lead
to a morphological bifurcation of the drop which takes the shape of a
star \cite{Holter52,Takaki84,Takaki85,Strier00,Aranson08}. Similar
star-shaped drops have been reported in drops vertically vibrated on
non-sticky surfaces, and the shape is generally attributed to a parametric
instability \cite{Yoshiyasu96}. Our original question was whether
oscillations could perhaps be explained even in the limit of viscous
drops, which we focus on in this paper. This is not the case, since
both our asymptotic results and simulations of the complete dynamics 
(i.e. beyond linear stability analysis) show that once unstable, a drop breaks up owing to the formation of a chimney.

We treat both the liquid drop and the surrounding gas
in the inertialess (Stokes) limit. For the asymptotic analysis, we
also require the drop to be much more viscous than the gas. The main
effect of this assumption is that there is hardly any flow inside the
drop, so it can be treated as being in hydrostatic equilibrium at
any instant in time. We also prescribe the rate at which gas is injected
into the underside of the drop, thus ignoring the possible interplay
between drop dynamics and vapor production in the Leidenfrost problem.

Our analysis is similar in spirit to the earlier paper of Duchemin et al.
\cite{Duchemin05}, but we only address the simpler case of a flat substrate.
As a result, we are able to perform all the calculations analytically
(up to a few universal constants, which have to be computed numerically).
Our solution curves are in qualitative agreement with those for a curved
substrate \cite{Duchemin05}, but now imply a full analytical description.
In addition, we are the first to perform a systematic stability analysis
of the stationary states. We find the maximum stable radius

\begin{equation}
\frac{r_{\rm max}}{\ell_c} \approx 4.35 - \tilde{r}
\end{equation}
where $\tilde{r}$ goes to zero in the limit of vanishing gas flow.

For typical experimental flow rates we find that $\tilde{r} \approx 0.4$,
consistent with the experimental result (\ref{eq.exp}).
At the end of the paper we discuss how our analysis relates to the
stability argument of \cite{These_Biance,Biance03}, based on the
Rayleigh-Taylor instability.

\section{Problem formulation}

\subsection{Geometry and dimensionless parameters}

\begin{figure}
\begin{center}
\includegraphics[width=8.0cm]{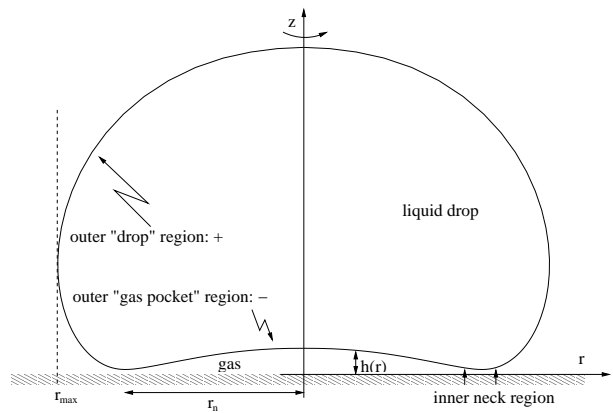}
\caption{Definitions and sketch of the matching regions.}
\label{fig:outer}
\end{center}
\end{figure}

We consider axisymmetric drops of liquid, levitated above a flat
surface by gas flowing into the underside of the drop, cf.
Fig.~\ref{fig:outer}. We set out to find the shape of stationary
drops and their stability, as a function of the gas flow rate and
the drop volume. The size of the drop is  expressed by the Bond number

\begin{eqnarray}
{\rm Bo} = \frac{R^2}{\ell_c^2},
\end{eqnarray}
where $V$ is the volume of the liquid drop, and $R=(4\pi V/3)^{1/3}$
the unperturbed radius. The dimensionless gas flow rate supporting
the drop reads

\begin{eqnarray}\label{eq:gamma}
\Gamma = \frac{Q \eta_{\rm gas}}{\ell_c^2 \gamma},
\end{eqnarray}
where $Q$ is the volume of gas that escapes through the narrow
neck region (see Fig.~\ref{fig:outer}) per unit of time, and
$\eta_{\rm gas}$ is the viscosity of the gas. Our
analysis will identify the flux $Q$ as the relevant quantity, which
can be calculated by integrating the gas flux entering from underneath
up to the neck position $r_n$. Let us also introduce a slightly different
dimensionalization of the flow rate

\begin{equation}
\chi = \frac{6 \Gamma \ell_c}{\pi r_n} = \frac{6 Q_n}{\pi r_n \ell_c},
\end{equation}
which will appear naturally in the analysis.

Finally, another parameter is the viscosity ratio between liquid
and gas
\begin{eqnarray}
\lambda = \frac{\eta_{\rm drop}}{\eta_{\rm gas}},
\end{eqnarray}
but which will be considered asymptotically large for most of this paper.
Throughout, lengths will be expressed in $\ell_c$, velocities in
$\gamma/\eta_{\rm gas}$, and stresses in $\gamma/\ell_c$.

\subsection{Structure of the problem}
The problem we set ourselves is to solve the inertialess, axisymmetric
fluid flow equations, with a prescribed influx of gas into the underside
of the drop. Most of our analytical work assumes in addition that the drop
is much more viscous than the gas. The structure of the expected solution
is shown in Fig.~\ref{fig:outer}. The gas pressure below the drop has to
be sufficiently large in order to support the weight of the drop. In
the limit of small dimensionless gas flux $\Gamma$, the gap between the
drop and the substrate must therefore be small in order to generate
enough pressure. The underside of the drop inflates to a gas pocket,
whose width is of similar size as the drop itself. The narrow gap is
formed in a small neck region only, where a large curvature assures
that the gas pressure can be sustained by corresponding surface tension
forces. Apart from this viscous neck region the gas pressure is constant,
both in the gas pocket as well as to the exterior of the neck.

This leads to the following asymptotic structure of the problem,
characterized by the matching between three different regions.
In the limit of small flux all viscous effects become localized in
a small neck region, situated at a radius $r = r_n$ from the center.
In this region, there exists a balance between viscous and surface
tension forces. In addition, the slope of the gap profile $h(r)$
turns out to be small in this region, so lubrication theory \cite{Batch67}
permits to reduce the flow equations to an ordinary differential equation
for $h(r)$. We will call this the inner solution or neck region.

To close the problem, boundary conditions are needed. These are provided
by two outer regions on either side of the neck, denoted by $'-'$ (the
gas pocket toward the center of the drop), and $'+'$ (the outside of the
drop). Both regions are controlled by a balance of gravity and surface
tension alone. First, we solve the equations in each of the regions
individually. Second, we require that both the slope and the curvature
of the profile match smoothly at the boundaries between two regions.
This leads to a set of equations which determines stationary drop solutions
uniquely. Solutions exist only below a certain critical neck position
$r_c$, in which case we find two branches, one with a small gap width
(the lower branch), and an upper branch with a larger gap width.

Our stability analysis of the two branches is based on the observation
that the relevant dynamic variable is the position $r_n$ of the neck,
which can shift easily. The maximum stable neck radius does {\it not}
coincide with $r_c$, but is significantly smaller, located on the lower
branch.

\section{Inner solution: neck region}

\subsection{Lubrication approximation}

We consider incompressible, axisymmetric flow in the gas layer,
so that mass conservation gives

\begin{equation}\label{eq:continuity}
r \dot{h} +
\left( r h \bar{u} \right)' = r v(r).
\end{equation}
Here $\bar{u}$ is the depth averaged horizontal velocity of the air
layer, while $v(r)$ is the rate at which air volume is injected per
unit area below the drop.
The main focus of the paper will be on stationary states and their
stability. Stationary drop profiles are found by taking $\dot{h}=0$,
and integrating (\ref{eq:continuity}) to

\begin{equation}\label{eq:steady}
rh \bar{u} = \frac{\Gamma(r)}{2\pi},
\end{equation}
where $\Gamma(r)=2\pi \int_0^r dr' r' v(r')$ is the flux in the lubrication
layer. In the case where the injection source is localized at $r=0$,
the flux $\Gamma$ is simply constant.

To get a closed equation for $h(r)$ in the neck region, we solve for
$\bar{u}$. As our results will confirm, the neck region is shallow,
$h' \ll 1$, meaning that we can use the lubrication approximation
\cite{Batch67} to analyze the flow, see Fig.~\ref{fig:outer}. Owing to
the large viscosity ratio between the drop and the surrounding gas,
the liquid drop acts as a no-slip boundary, and the flow in the gas
layer is well approximated by

\begin{equation}
u = 6 \bar{u} \left(\frac{z}{h} - \frac{z^2}{h^2} \right).
\label{eq:u}
\end{equation}
Since the Reynolds number is very small in typical experiments, we
use the Stokes approximation \cite{Batch67} to relate this velocity
to the pressure. As there is almost no flow inside the drop,
the liquid will be at hydrostatic equilibrium, $p_{\rm liquid}=p_0 - z$.
Furthermore, the pressure jump across the interface equals
the curvature times the surface tension, so one obtains the lubrication
pressure inside the gas layer as

\begin{equation}
p = p_0 -  h - h''.
\label{eq:pressure}
\end{equation}
In what follows we will show that the width of the neck region scales
as $\Gamma^{1/5}$ and thus is asymptotically small in the limit of
vanishing flux. We are therefore permitted to neglect the axisymmetric
contribution to the curvature in the neck region. Using the horizontal
component of the Stokes equation, $p'= \partial^2 u/\partial z^2$,
we find

\begin{equation}
\bar{u} = 12 h^2 \left( h' + h''' \right).
\label{baru}
\end{equation}
Now (\ref{eq:steady}),(\ref{baru}) provide a closed equation for the
stationary interface profile $h(r)$:

\begin{equation}\label{eq:steady1}
h^3\left( h' + h''' \right) = \frac{6\Gamma(r)}{\pi r}.
\end{equation}
The right hand side of (\ref{eq:steady1}) represents the viscous stress
in the flow, and will only become important when $h$ is small, i.e. in a
small neck region around $r_n$, where we may set $r = r_n$. This gives

\begin{equation}\label{eq:steady2}
h^3\left( h' + h''' \right) = \chi ,
\end{equation}
with

\begin{equation}
\chi \equiv \frac{6\Gamma(r_n)}{\pi r_n}.
\end{equation}

A crucial observation is that there is no need to know the precise
form of how the gas is injected, but one only requires the flux
across the neck. This of course provides a great simplification for
the Leidenfrost problem, where evaporation rates are related in a
complicated way to the temperature profile inside the drop.

\subsection{Similarity solution for neck region}

As gravity is unimportant in the thin neck region, (\ref{eq:steady2})
can be further simplified to

\begin{equation}\label{eq:inner0}
h^3 h''' = \chi.
\end{equation}
Since we are interested in the limit of small flux, we look for
similarity solutions

\begin{equation}
h(r) = \chi^\alpha H\left( \xi \right),
\quad {\rm where} \quad \xi= \frac{r-r_n}{\chi^\beta}
\end{equation}
giving

\begin{equation}\label{eq:inner}
H^3 H'''=1, \quad {\rm with} \quad 4\alpha - 3\beta = 1.
\end{equation}
In the limit $\xi \rightarrow \infty$, the solutions have to match
onto a sessile drop of constant curvature. Since

\begin{equation}
h'' = \chi^{\alpha - 2\beta} H'' ,
\end{equation}
one needs that $\alpha - 2\beta = 0$ for the curvature to remain
finite as $\chi\rightarrow 0$. Together with (\ref{eq:inner}) this
fixes $\alpha=2/5$ and $\beta=1/5$, and hence we have
\begin{equation}
h(r) = \chi^{2/5} H\left(\frac{r-r_n}{\chi^{1/5}} \right).
\label{hscal}
\end{equation}

The form of the similarity function will be determined
from the matching below. The fact that $\alpha > \beta$ justifies the
assumptions made so far. First, we find that $h'\ll 1$ in the limit
$\chi \rightarrow 0$, justifying the use of lubrication theory.
Similarly, $h' \ll h'''$, so that both gravity and the axisymmetric
curvature can indeed be neglected in the neck region.

The asymptotic behavior of (\ref{eq:inner}) is quadratic for both
$\xi \rightarrow \pm \infty$,

\begin{eqnarray}
H_+ &=&
\frac{1}{2} K_+ \xi^2 + S_+ \xi \quad {\rm for}
\quad \xi \rightarrow \label{eq:plus} \infty\\
H_- &=&
\frac{1}{2} K_- \xi^2 + S_- \xi \quad {\rm for}
\quad \xi \rightarrow -\infty. \label{eq:minus}
\end{eqnarray}
Physically, the values of the asymptotic curvatures $K_\pm$ set the pressure
in the corresponding outer regions.

\begin{figure}
\begin{center}
\includegraphics[width=8.0cm]{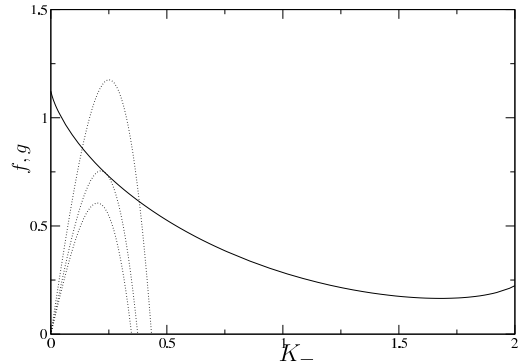}
\caption{Solid line: the function $f$ relates the slope $S_-$ to
curvature $K_-'$ of the inner solution (Eq.~(\ref{eq:fscal}) with $K_+=2.17$,
corresponding to $r_n=r_0$). Dotted lines: the function $g$ provides the
matching condition between inner region and gas pocket region
(Eq.~(\ref{eq.gexp}) with $\chi=10^{-7}$). The three curves correspond
to values $r_n=3.55$ (below critical), $r_n=3.62$ (critical), and
$r_n=3.65$ (above critical).
               }
\label{fig:f}
\end{center}
\end{figure}

Since (\ref{eq:inner}) is of third order, solutions can be specified
by three independent parameters, one of which can be absorbed into a
shift of $\xi$. Therefore, the two asymptotic curvatures $K_\pm$ uniquely
determine the solution. As a consequence, the slopes $S_\pm$ are dependent
variables. To perform the matching, we require the function

\begin{equation}
S_- =  - f \left( K_-, K_+ \right), \label{eq:f}
\end{equation}
whose existence is assured by the above argument.
Since (\ref{eq:inner}) is invariant under the transformation
$h\rightarrow h/a$ and $x\rightarrow x/a^{4/3}$ one must have

\begin{equation}
f \left( K_-, K_+ \right) = K_+^{1/5} f \left( \frac{K_-}{K_+},1 \right).
\label{eq:fscal}
\end{equation}
This function is computed numerically and is plotted in Fig.~\ref{fig:f}.
We show below that stationary solutions correspond to the intersection
of $f$ with another function $g$, shown in the same figure. It can be
seen that the matching breaks down at a critical neck radius $r_n$,
beyond which stationary solutions cease to exist.

\section{Outer solutions}\label{sec.outer}

Having seen that viscous effects are localized in the neck region,
the rest of the drop is at static equilibrium. Hence, the pressure
is constant both in the gas pocket between the drop and the substrate,
as well as to the exterior of the neck. These pressures are not equal,
however, since one requires a pressure difference to drive the flow
across the neck. In Fig.~\ref{fig:outer} we therefore distinguish
two outer regions, denoted by + and - respectively. Since
$p_\pm = p_0 -  h - \kappa$, the outer solutions can be obtained from

\begin{equation}\label{eq.static}
\kappa + h = c_\pm,
\end{equation}
where $\kappa$ is the curvature of the interface. The constants
$c_\pm$ determine the pressure difference across the neck

\begin{equation}
\Delta p=p_- - p_+ = c_+ - c_-,
\label{dp_neck}
\end{equation}
and will follow from the matching.

\subsection{Outer "drop" region: +}

\begin{figure}
\begin{center}
\includegraphics[width=8.0cm]{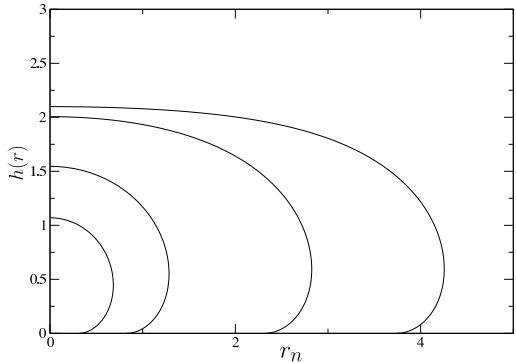}
\caption{The outer solution of the "drop" region corresponds to a
perfectly non-wetting sessile drop. The size of the drop, characterized
by $r_n$, sets the curvature $K_+$ for the inner solution.}
\label{fig:drops}
\end{center}
\end{figure}

Below we will find that the profile of the "drop" region requires
$dh/dr\rightarrow 0$ as $h\rightarrow 0$, in order to match to the
neck smoothly. This corresponds to a perfectly non-wetting sessile
drop (Fig.~\ref{fig:drops}). When matching the curvature we also
require $d^2h/dr^2=K_+$ as $h\rightarrow 0$. Owing to the vanishing
slope near $h=0$ we are allowed to write $\kappa=d^2h/dr^2$ in
(\ref{eq.static}). Hence, one finds $c_+=K_+$.

To deal with the overhang of the sessile drop, it is convenient to
solve the profile in terms of the arclength $s$ along the interface.
We define $\theta$ as the angle with the horizontal and rewrite
(\ref{eq.static}) as

\begin{eqnarray}
\frac{d\theta}{ds} &=& - \frac{\sin \theta}{r} - h + K_+ \\
\frac{dh}{ds} &=& \sin \theta \\
\frac{dr}{ds} &=& \cos \theta,
\end{eqnarray}
with boundary conditions

\begin{eqnarray}
\theta(0) &=& 0 \\
h(0) &=& 0 \\
r(0) &=& r_n \\
\theta(s_t) &=& \pi \\
r(s_t) &=& 0,
\end{eqnarray}
where $s_t$ is the value of the arclength at the top. Two of these
five boundary conditions serve as the definitions of $r_n$ and $s_t$,
so that the remaining three boundary conditions fix the solution uniquely.
The equations have been solved numerically.

\begin{figure}
\begin{center}
\includegraphics[width=8.0cm]{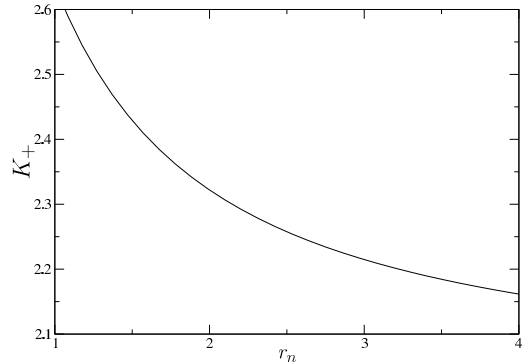}
\caption{The outer solution fixes the value of $K_+$ as a function of
the neck radius (which controls the drop volume). The maximal
radius $r_0=3.38317\cdots$ gives $K_+ = 2.17\cdots$.
}
\label{fig:kplus}
\end{center}
\end{figure}

Each value of $K_+$ thus gives a solution with a different $r_n$, some
of which are shown in Fig.~\ref{fig:drops}. The numerically obtained
relation between $K_+$ and $r_n$ is depicted in Fig.~\ref{fig:kplus}.
For the {\it maximal} neck radius $r_n = r_0 = 3.38317\cdots$, introduced
below, one finds $K_+ = 2.17\cdots$.

The value of $r_n$ effectively sets the volume of the drop.
Namely, the weight of the sessile drop is carried by the pressure
exerted by the substrate on the contact area $\pi r_n^2$. The difference
between the liquid and the gas pressures at $h=0$ is simply $K_+$, so we find

\begin{equation}\label{eq.vplus}
K_+ \pi r_n^2 = 2\pi V_+  \quad \Rightarrow \quad
V_+ = \frac{1}{2}K_+ r_n^2.
\end{equation}
where $V_+$ is defined as the real volume divided by $2\pi$, i.e.

\begin{equation}
V_+ = \frac{1}{2\pi} \int_0^{h_{\rm max}} dh \,\pi r^2.
\end{equation}
Note that to obtain the real liquid volume, one has to subtract the volume
$V_-$ of the gas pocket. However, $V_-$ goes to zero in the limit of
vanishing gas flow, as shown below.

\subsection{Outer "gas pocket" region: - }

In the "gas pocket" region, the profile $h(r)$ is no longer multivalued
and we can express the curvature as

\begin{equation}
\kappa = \frac{h''}{(1+h'^2)^{3/2}} + \frac{h'}{r(1+h'^2)^{1/2}}.
\end{equation}
The solution is then specified by (\ref{eq.static}) with boundary conditions

\begin{eqnarray}
h'(0) &=& 0 \\
h(r_n) &=& 0.
\end{eqnarray}
In Appendix~\ref{app.bessel} we show that the solution can be written as an
expansion

\begin{equation}
h(r) =  -c_- \frac{ J_0(r) - J_0(r_n)}{J_0(r_n)} +
\mathcal{O}\left( c_-^3\right),
\end{equation}
where $J_0(r)$ is a Bessel function of the first kind.
Using furthermore that the curvature has to match the curvature of
the inner solution $K_-$, and thus $K_- = h''(r_n)$,
we can further simplify to

\begin{equation}\label{eq:bessel}
h(r) =  K_- \frac{ J_0(r) - J_0(r_n)}{J_0''(r_n)} +
\mathcal{O}\left( K_-^3\right).
\end{equation}
We see that the thickness scale of the gas pocket is set by
the value of $K_-$. In the limit of vanishing flux we expect this
thickness to tend to zero, making $K_-$ a small parameter.
To find solution branches, it is crucial to go beyond linear order
and to find the term of order $K_-^3$ in (\ref{eq:bessel}). The only
quantity that is needed to perform the matching to (\ref{eq:f}),
coming from the inner solution, is the slope $h_-'(r_n)$. This calculation
is done in Appendix~\ref{app.bessel}.

\begin{figure}
\begin{center}
\includegraphics[width=8.0cm]{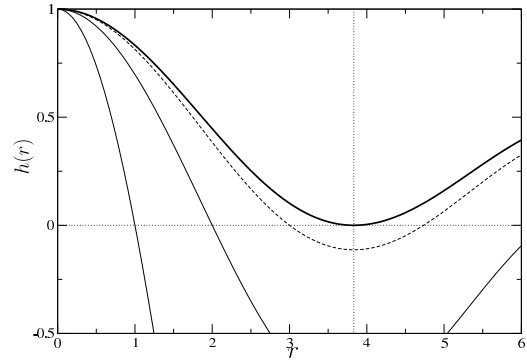}
\caption{
Outer solutions for the gas pocket region (amplitudes normalized to
unity). Thin solid lines correspond to $r_n=1,2$; the dashed line,
corresponding to $r_n=3$, illustrates that $h$ would have to become
negative to realize a neck radius larger than $r_0$. The heavy line shows
the maximum possible $r_n=r_0$, corresponding to the first minimum of
$J_0(r)$.
     }
\label{fig:pocket}
\end{center}
\end{figure}

At this point we can already infer an upper bound on the possible
values of $r_n$. Figure~\ref{fig:pocket} shows the outer gas pocket solution
(with normalized amplitude) for various values of $r_n$. The outer
solutions are defined on the domain where $h(r)\geq 0$, hence the maximum
possible neck radius is achieved when $J_0(r)$ has its first minimum
at the maximal radius $r_0=3.8317\cdots$ (vertical line). The
corresponding solution is drawn with a heavy line.

A second remark is that $J_0''$ vanishes at $r\approx 1.852\cdots$,
so $K_-$ must become zero at this radius. At even smaller radii $J_0''$
turns negative, which yields negative values of $K_-$. However, the inner
solution cannot reach large $h$ for negative $K_-$, which means
that the matching procedure described here does not work.
Dealing with this problem requires an additional matching region between
the inner and outer (gas pocket) solution, the introduction of which
lies beyond the scope of this paper. We will simply stay away from
$r_n \approx 1.852$ and instead focus on radii close to the maximal
value $r_0 \approx 3.8317$, as detailed below.

\section{Matching the asymptotic regions}

\subsection{Matching conditions}

We can now match the asymptotic regions by expressing
(\ref{eq:plus}),(\ref{eq:minus}) in their original variables and
expanding the outer solutions around $r=r_n$,

\begin{eqnarray}
h_{\rm out \pm} &=& \frac{1}{2} h_\pm '' |
 _{r_n} (r-r_n)^2 + h_\pm'|_{r_n} (r-r_n) \\
h_{\rm in \pm} &=& \frac{1}{2} K_{\pm} (r-r_n)^2 + \chi^{1/5} S_{\pm} (r-r_n).
\end{eqnarray}
Therefore the matching conditions become

\begin{eqnarray}
K_{\pm} &=& h_\pm'' |_{r_n} \\
\chi^{1/5} S_{\pm}' &=&  h_\pm' |_{r_n}.
\end{eqnarray}

The conditions on the curvature were already taken into account when
computing the outer profiles from (\ref{eq.static}). Typical values
for $K_+$ are of order unity, while the slope requires $h_+' |_{r_n}=0$
as $\chi \rightarrow 0$. This is why for the first outer solution we
considered a perfectly non-wetting drop.

The $-$ conditions are more subtle. The thickness of the gas pocket
goes to zero asymptotically so that both $h_-'' |_{r_n}$ and
$h_-' |_{r_n}$ will be small. In this case the selection of the solution
explicitly requires the slope condition, which we express as

\begin{equation}\label{eq:g}
S_- =  \frac{K_-}{\chi^{1/5}} \frac{h_-' |_{r_n}}{h_-'' |_{r_n}}
\equiv - g(K_-,r_n;\chi).
\end{equation}
Together with (\ref{eq:f}) this closes the matching problem:

\begin{eqnarray}
f\left (K_-, K_+\right) = g \left(K_-,r_n ; \chi \right).
\label{eq:match}
\end{eqnarray}
This equation indeed contains the three matching regions: $K_+$
implicitly depends on $r_n$ through the $+$ outer solution,
$f$ is determined by the inner solution, while $g$ follows from
the $-$ outer solution.

\subsection{Bifurcation: critical radius $r_c$}

\begin{figure}
\begin{center}
\includegraphics[width=8.0cm]{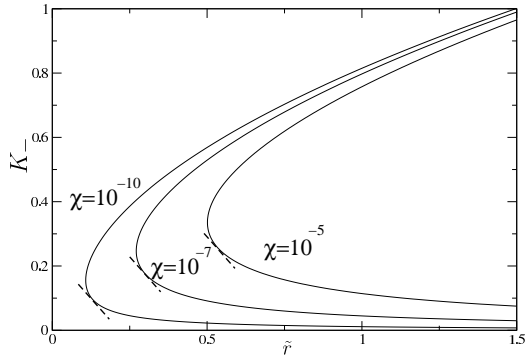}
\caption{The bifurcation diagram ($K_-,\tilde{r}$), derived from
(\ref{eq:bifurc}). Curves correspond to different values of the flux,
$\chi=10^{-5},10^{-7},10^{-10}$, revealing the weak dependence on $\chi$.
The dashed lines represent perturbations
$\delta \tilde{r}, \delta K_-=-\delta \tilde{r}/c$ which are tangent to the
solution curve. They represent marginal perturbations, separating
stable from unstable solutions.
     }
\label{fig:bifurcation}
\end{center}
\end{figure}

For a given value of the flux $\chi$, we have reduced the problem to
finding the intersections of the functions $f$ and $g$. This is
sketched in Fig.~\ref{fig:f}, showing $f$ and $g$ for $\chi=10^{-7}$
and several values of $r_n$. Depending on the value of $r_n$, there
can be two intersections, one intersection (when the curves are tangent),
or no intersection. Each intersection corresponds to a stationary
drop solution. This can be translated into a bifurcation diagram
showing $K_-$ vs $r_n$ (Fig.~\ref{fig:bifurcation}). For small radii
there are two branches of solutions, corresponding to the two
intersections, which merge at $r_c$. No stationary drop solutions
exist for $r_n> r_c$.

\begin{figure}
\begin{center}
\includegraphics[width=8.0cm]{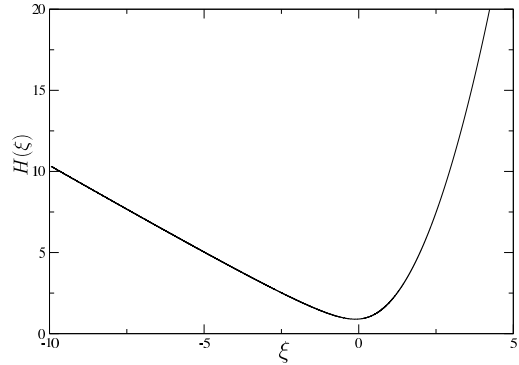}
\caption{The inner solution $H(\xi)$ obtained from numerical integration of
(\ref{eq:inner}), with $K_+=2.17$ and $K_-=0$. It will follow that these
values correspond to the critical solution. The minimum value $H_n \approx
0.931$ determines the thickness of the neck (\ref{eq.hn}).
                   }
\label{fig:inner}
\end{center}
\end{figure}

We analyze the bifurcation in the limit of vanishing flux,
$\chi \rightarrow 0$. We will show that

\begin{equation}
r_c = r_0 + \mathcal{O}\left(\chi^{2/15} \right),
\end{equation}
so that the critical neck radius $r_c$ approaches the maximal
radius $r_0$ in the limit of vanishing flux.
To analyze the vicinity of the critical point we introduce
\begin{equation}
\tilde{r} = r_0 - r_n.
\end{equation}
At the same time we will find that $K_- \propto \chi^{1/15}$.
This means that as the limit of $\chi$ going to zero is reached,
$K_- = 0$ and $r_n = r_0$, which implies $K_+=2.17$ according to
Fig.~\ref{fig:kplus}. These two data fix the solution of
(\ref{eq:inner}) uniquely, and lead to the asymptotic profile
shown in Fig.~\ref{fig:inner}. From its minimum, one finds that
\begin{equation}
h_n \approx 0.931 \,\chi^{2/5},
\label{eq.hn}
\end{equation}
in agreement with the scaling found by \cite{Duchemin05}.

We now analyze the first correction to the solution as $\chi$ increases,
but in the limit where $\chi, K_-, \tilde{r} \ll 1$. This can be
done by considering the corresponding limit of the functions $f$ and $g$,
cf. Fig.~\ref{fig:f}.
Namely, the function $f$ approaches a constant, which is found
numerically to be
\begin{eqnarray}
f \simeq f_0 = 1.12 \cdots .
\end{eqnarray}
on the other hand, the asymptotic form of $g$ becomes

\begin{equation}\label{eq.gexp}
g \simeq \chi^{-1/5}  K_-\left( \tilde{r} -  g_2 K_-^2  \right) .
\end{equation}
The first term of (\ref{eq.gexp}) is found by expanding (\ref{eq:bessel})
for $r_n$ close to $r_0$:

\begin{eqnarray}
\frac{h_-' |_{r_n}}{h_-'' |_{r_n}} &=&   \frac{J_0''(r_0)(r-r_n)}{J_0''(r_n)}
+ \mathcal{O}\left( K_-^2 \right) \nonumber \\
&\simeq& - \tilde{r} + \mathcal{O}\left( K_-^2 \right),
\end{eqnarray}
where we used the property $J_0'(r_0)=0$. We need to keep the $K_-^2$
term as it can become of the same order as $\tilde{r}$. For details
we refer to Appendix~\ref{app.bessel}, where we show that $g_2=1.486\cdots$.

The matching condition $f=g$ (cf. (\ref{eq:match})) is now reduced to
a horizontal line intersecting a cubic function:
\begin{equation}
f_0 = \frac{K_-}{\chi^{1/5}}(\tilde{r} - g_2 K_-^2).
\label{eq:matchs}
\end{equation}
Solving for $\tilde{r}$, one finds

\begin{equation}\label{eq:bifurc}
\tilde{r}(K_-,\chi) = \frac{\chi^{1/5}f_0}{K_-} + g_2 K_-^2,
\end{equation}
which has been plotted for different values of $\chi$ in
Fig.~\ref{fig:bifurcation}. Thus for a given value of $r_n$
one finds {\it two} solution branches, which end at the critical
value

\begin{equation}\label{eq.rc}
\tilde{r}_c = 3  \left(\frac{1}{4} g_2 f_0^2 \right)^{1/3} \chi^{2/15} +
\mathcal{O}\left( \chi^{4/15}\right),
\end{equation}
as claimed before. Note that the smallness of the power $2/15$ makes
$\tilde{r}_c$ non-negligible. For typical experimental values of the flux
the critical point is thus substantially shifted with respect to
the asymptotic value $r_0$.

Plugging this back into (\ref{eq:bifurc}), one finds the
value of $K_-$ at the critical point:
$K_-^{(c)} = 0.72 \chi^{1/15}$. But it follows from (\ref{eq:bessel})
that $h_0 = h(0) \approx K_-(1-J_0(r_0))/J_0''(r_0)$, and thus the
maximum gap width is to leading order
\begin{equation}
h_0 \approx 2.52 \,\chi^{1/15}.
\label{eq.h0}
\end{equation}
This concludes the analysis of the stationary solutions, which are
described by (\ref{eq:bifurc}). At a given flow rate $\chi$, the
critical neck radius is given by (\ref{eq.rc}), which approaches
the maximal value $r_0$ in the limit $\chi\rightarrow 0$.

\section{Stability}

We now turn to the important question of which part of the solution
branches shown in Fig.~\ref{fig:bifurcation} are stable.
Essentially, we find that the lower branch is linearly stable, while the
upper is linearly unstable. Surprisingly, however, the marginal
point is not exactly at the maximum radius, but slightly before.

\subsection{Stability limit}

Once more we make use of the fact that there is no flow inside the
drop, so that the drop shape adjusts quasi-statically to variations
in the neck region. We therefore consider infinitesimal variations in
the neck position, $\delta r_n$ and assess the corresponding change in
levitation force $\delta F$. Since the pressure difference
$\Delta p=p_- - p_+$ across the neck acts for $r< r_n$, this force reads

\begin{equation}
F = \Delta p \pi r_n^2.
\end{equation}

A marginal perturbation $\delta r_n$ occurs whenever the resulting
levitation force is unchanged, $\delta F=F' \delta r_n=0$, so that it
still equilibrates the weight of the drop. Hence, we find the marginal
condition

\begin{equation}
\Delta p'= - \frac{2 \Delta p}{r_n},
\end{equation}
where the prime denotes the derivative with respect to $r_n$.
In order to produce the same levitation force, an increase in $r_n$ thus
has to be compensated by a decrease of  $\Delta p$. Had the pressure
stayed constant, $F$ would be larger than the weight of the drop leading
to the formation of a chimney and thus to instability. Similarly,
pressures smaller than the marginal condition leads to a stable
situation, giving the stability criterion

\begin{equation}
\Delta p'  - \frac{2 \Delta p}{r_n} < 0.
\end{equation}
In the limit of small $\chi$, the pressure difference (\ref{dp_neck})
is simply the difference of the curvatures:

\begin{equation}
\Delta p = K_+ - K_-,
\end{equation}
so that stability requires

\begin{equation}
K_+' - K_-' + \frac{2 \Delta p}{r_n} < 0.
\end{equation}
The derivative $K_+'$ can be read off from Fig.~\ref{fig:kplus}, and
is negative. Clearly, this has a stabilizing effect. The sign of
$K_-'$ can be inferred from the bifurcation diagram. The lower branch
has a stabilizing contribution, while the upper branch is be destabilizing.
The location of the marginal point, however, depends on the numerical
values of the three terms.

Taking the derivative of (\ref{eq.vplus}), we find

\begin{equation}
K_+' = \frac{2V_+'}{r_n^2} - \frac{2K_+}{r_n}.
\end{equation}
Moreover, for vanishing flux $K_- \ll K_+$, hence we may replace
$\Delta p \simeq K_+$, giving the stability criterion

\begin{equation}\label{eq.stability}
K_-' > \frac{2V_+'}{r_n^2}.
\end{equation}
Near the maximal radius $r_n\approx r_0$ the criterion
for stability becomes
\begin{equation}\label{eq.stability_asymp}
K_-' > c^{-1} \equiv \frac{2V_+'}{r_0^2} = 0.92\cdots.
\end{equation}
Indeed, the upper branch with $K_-'<0$ is unstable, but the marginal
point is not at the maximum radius $K_-'=0$, but slightly before. 
This is indicated in Fig.~\ref{fig:bifurcation} by the dashed lines, 
which each have a slope $0.92$.

\subsection{Linear stability analysis}

We now include dynamics into the stability analysis. We note first
that an infinitesimal variation of the neck position,
$\delta r_n=-\delta \tilde{r}$, also induces a variation of the curvature,
$\delta K_-$, and of the flux, $\delta \chi$. These three parameters
are related through mass conservation of the liquid and the gas.
The analysis is closed by a third equation coming from matching the
dynamic inner region to the hydrostatic outer regions.

The volume of the liquid can be calculated by the volume enclosed by
the sessile drop solution, $V_+$, minus the volume of the gas pocket
$V_-$, i.e.

\begin{equation}
V_{\rm liquid} = V_+(r_n) - V_-(r_n,K_-).
\end{equation}
This is exact up to asymptotically small corrections due to the inner
region. The volume $V_+$ is (numerically) determined by the value of
$r_n$, while $V_-$ can be computed analytically using  (\ref{eq:bessel}),

\begin{eqnarray}
V_-(r_n,K_-) = \int_0^{r_n} dr \,r\, h_-(r) \nonumber \\
\simeq \frac{1}{2} K_- r_n^2 \left(
\frac{J_0(r_n) - 2 J_1(r_n)/r_n }{J_0(r_n) - J_1(r_n)/r_n}
\right).
\end{eqnarray}
The expression simplifies at $r_0$, because of the property
$J_1(r_0)=0$. Since the liquid volume is strictly conserved,
$\dot{V}_{\rm liquid}=0$, one finds near $r_0$

\begin{equation}\label{eq:hups}
\delta K_- = - \frac{ \delta \tilde{r}}{c},
\end{equation}
where the constant $c$ has been defined by (\ref{eq.stability_asymp}).
Relation (\ref{eq:hups}) expresses the fact that when $r_n$ increases,
increasing $V_+$, the volume of the gas pocket has to increase by a
similar amount to keep the liquid volume constant. This is achieved
by an increase of $K_-$.

Mass conservation of the gas is described by continuity
(\ref{eq:continuity}), which can be integrated to

\begin{equation}
 r h \bar{u} = \Gamma(r) - \frac{\partial}{\partial t} \int_0^{r} dr \,r \,h .
\end{equation}
The second term on the right hand side can be identified as the rate of
change of gas pocket volume $\dot{V}_-$, which we will write as
$-V_+' \delta \dot{\tilde{r}}$. This change absorbs part of the injected
air, decreasing the flux passing across the neck. Considering the
radius somewhere inside the neck region, $r\approx r_n$, the equation
can be simplified to (using (\ref{baru}) and $h'\ll 1$):

\begin{equation}\label{eq:innerdyn}
h^3 h''' = \chi + \delta \chi,
\end{equation}
where the variation of the flux reads

\begin{equation}\label{eq:deltachi}
\delta \chi = \frac{r_0}{24c} \delta \dot{\tilde{r}}.
\end{equation}

The matching condition (\ref{eq:bifurc}) closes the dynamical
system, taking into account the dependencies (\ref{eq:hups}) and
(\ref{eq:deltachi}).
The marginal case $\delta\chi = 0$ corresponds to a curve tangent
to any of the lines $\tilde{r}(K_-)$ shown in Fig.~\ref{fig:bifurcation}.
Since in addition the slope of such a tangent curve must be $-c^{-1}$
according to (\ref{eq:hups}), this uniquely fixes a point on any
of the lines at constant $\chi$. The critical tangent curve is drawn
dashed in Fig.~\ref{fig:bifurcation}. Below this point, on the lower branch,
solutions are stable, above they are unstable.

Formally, the growth rate of perturbations is computed by writing

\begin{equation}
\delta \dot{\tilde{r}} = \sigma \delta \tilde{r}.
\end{equation}
Now using (\ref{eq:hups}), (\ref{eq:deltachi}), and the
first variation of (\ref{eq:bifurc}) one finds

\begin{equation}
\sigma = \frac{24}{cr_0}\left(
\frac{\partial \tilde{r}}{\partial \chi} \right)^{-1}
\left[
\frac{\partial \tilde{r}}{\partial K_-} + c
\right].
\end{equation}
The partial derivatives are to be evaluated from (\ref{eq:bifurc}).

This indeed gives the same stability boundary as (\ref{eq.stability}),
which was based on a global force balance (note that
$\partial \tilde{r}/\partial K_- = - (K_-')^{-1}$).
The maximum stable radius $\tilde{r}_s$ is found by the condition $\sigma=0$,
yielding

\begin{equation}
\frac{\chi^{1/5} f_0}{K_-^2} - 2g_2K_- = c
\label{inter}
\end{equation}
as an equation for $K_-$ at the stability boundary. This value of
$K_-$ is inconsistent with the asymptotic estimate
$K_-\approx \chi^{1/15}$ considered so far, indicating that the point where the
solution exchanges stability is at a distance slightly larger from
the critical point $r_c$. This means that $K_-$ is {\it smaller} than
expected (further down the lower branch, cf. Fig.~\ref{fig:bifurcation}).
Thus the second term on the right of (\ref{inter}) is small compared
to the other two, and we obtain

\begin{equation}
K_- = \left(\frac{f_0}{c}\right)^{1/2}\chi^{1/10}.
\label{inter2}
\end{equation}
If we evaluate (\ref{eq:bifurc}) in the same limit, we finally obtain

\begin{equation}\label{eq:hopla}
\tilde{r}_s = (f_0 c)^{1/2}\chi^{1/10}.
\end{equation}
Thus for vanishing flux the maximum radius of stable
solutions approaches $r_0$, but with an even smaller power than
$r_c$. This scaling implies that $r_s < r_c < r_0$, as seen
in Fig.~\ref{fig:bifurcation}.

\section{Numerical tests}

\subsection{Non-linear dynamical behavior }

We begin with a simulation of the full axisymmetric Stokes problem,
using a boundary integral method \cite{RA78}, which has the advantage that it
tracks the interface with high precision. The idea is to regard the
interface as a continuous distribution of point forces, which point
in the direction of the normal and whose strength is proportional to the
mean curvature. Since for Stokes flow one knows the Green function
giving the velocity field resulting from a point force, one can write
the velocity anywhere in space as an integral over the free surface.
In an axisymmetric situation, the angle integral can be performed,
so the remaining integration is one-dimensional.

External flow sources can simply be added; in the present case
we take the gas flow as a point source of strength $Q$ situated
at the origin on the solid plate which bounds the flow. For this
a simple exact solution is available \cite{HB83}. Likewise for the
Green function one must take into account the presence of a no-slip
wall. This is possible using the method of images \cite{Blake71},
and the resulting boundary integral formulation has been applied
successfully to the motion of drops relative to a wall \cite{P90}. If, as in
our case, the viscosity of the drop is different from that of the
surrounding, one must account for the stress mismatch across the interface.
This can be done at the cost of introducing another integral over
the velocity on the interface into the equation, which turns the
equation for the velocity field into an integral equation. After solving
this equation for a given interface shape, the thus computed velocity
field can be used to advance the interface.

We follow closely an earlier implementation of the boundary integral
method, used for example the coalescence of two drops inside another
fluid \cite{ELS99}. The only significant difference is that the free space
Green function has been replaced by those for half space, bounded by
a wall. We tested the code by comparing to an exact solution of a sphere
moving perpendicular to a wall \cite{B68}. This is realized in the limit
of a very small drop, or of very large drop viscosity, so that there
is hardly any deformation. The agreement was good, but significant
deviations occurred when the gap between the wall and the drop was
smaller than 5 \% of the drop radius. At present, we do not know the
origin of this numerical problem, which prohibits us from investigating
the asymptotic limit of very small gap spacings. Instead, we report
on simulations at moderate gap spacings, which show the nonlinear stages
of chimney formation, not captured by our linear stability analysis.

\begin{figure}
\begin{center}
\includegraphics[width=8.0cm]{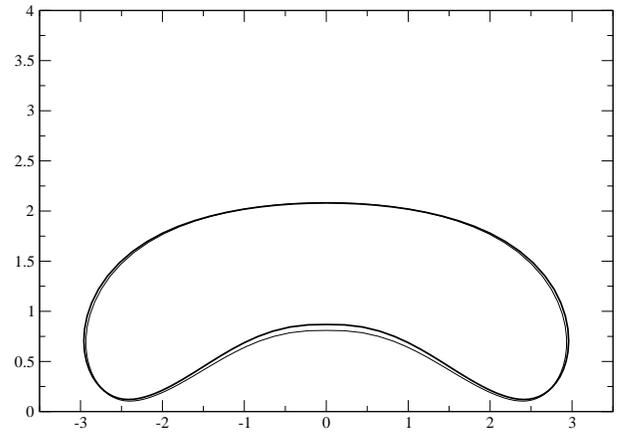}
\caption{Boundary integral simulation of a drop
with parameters $\Gamma = 0.02$, Bo=4.2, and $\lambda=100$.
The drop relaxes toward a stable state, which is drawn as the
heavy line.
                 }
\label{fig:stable}
\end{center}
\end{figure}
\begin{figure}
\begin{center}
\includegraphics[width=8.0cm]{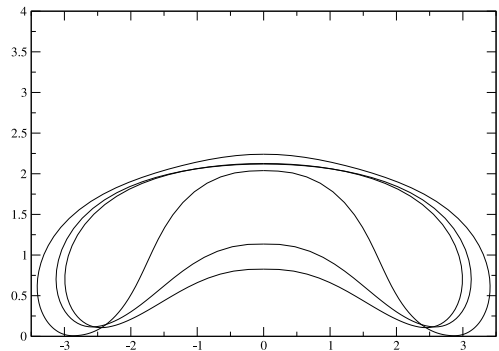}
\caption{The same as Fig.~\ref{fig:stable}, but with a slightly
larger Bo = 4.4. The air bubble under the center of the drop
lifts up to form a chimney. The time interval between the profiles
is $\Delta t$ = 3000, in units of $\ell_c \eta_{\rm gas} /\gamma$.
              }
\label{fig:chimney}
\end{center}
\end{figure}

Figure \ref{fig:stable} shows a viscous drop which is slightly
smaller then the stability boundary. Starting from a configuration
shown as the light curve, it relaxes toward a stationary stable state
(heavy line). For a Bond number which is just slightly larger,
the same initial condition leads to a rising gas bubble in center of
the drop, see Fig.~\ref{fig:chimney}. A thin film forms between the
rising gas bubble and the top of the drop, which drains slowly.
As seen in Fig.~\ref{fig:stable}, the neck radius is
$r_n\approx 2.5$, giving $\chi = 0.015$. On the basis of our
asymptotic theory (\ref{eq:hopla}), a rough estimate of the stability boundary 
gives $r_s\approx 3.2$.

\subsection{Lubrication approximation}

To test the bifurcation scenario in more detail we resort
to direct numerical simulation of the lubrication equation.
Due to the overhang of the drop, we separate the upper part of the drop and
the lower part of the drop at the maximum radius, $r_{\rm max}$, defined
by the point $|h'|=\infty$. The upper part is solved as described
in Sec.~\ref{sec.outer} A, and for the lower part of the
drop we use

\begin{eqnarray}
\kappa &=& \frac{h''}{(1+h'^2)^{3/2}} + \frac{h'}{r(1+h'^2)^{1/2}}\\
\chi &=& h^3 \left( \kappa' +h' \right).\label{eq:philippe}
\end{eqnarray}
This describes both the inner and outer regions in the lower part of
the interface, while we have conveniently taken the rate of injection
$\Gamma(r)/r$ to be constant for all $r$. Boundary conditions for this
3rd order equation are

\begin{eqnarray}
h'(0) &=& 0 \\
h'(r_{\rm max}) &=& \infty \\
\kappa (r_{\rm max}) &=& \kappa_{\rm patch},
\end{eqnarray}
where $\kappa_{\rm patch}$ is the curvature at the point where the
upper and lower solutions are patched. A one-parameter family of
solutions is obtained through variation of the upper part of the drop.
It was shown in \cite{Duchemin05} that this procedure provides
drop solutions that are quantitatively accurate.

\begin{figure}
\begin{center}
\includegraphics[width=8.0cm]{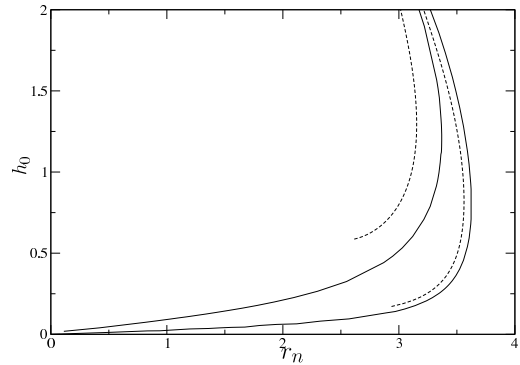}
\caption{Bifurcation diagram $h_0$ vs $r_n$ for $\chi=10^{-4}$ and $10^{-7}$.
Smaller $\chi$ yield larger radii. Solid lines were obtained from
numerical solution of the lubrication equation (\ref{eq:philippe}). Dashed
lines correspond to asymptotic theory (\ref{eq:bifurc}).}
\label{fig.lubri1}
\end{center}
\end{figure}

The numerically obtained drop profiles are conveniently characterized by the
position of the neck, $r_n$, and the gap below the center of the drop,
$h_0$. Numerical results for the solution branches are shown as solid lines in
Fig.~\ref{fig.lubri1} for two values of the flux. $\chi=10^{-4}$ is a
typical experimental value encountered for Leidenfrost drops, while
$\chi=10^{-7}$ illustrates the convergence toward the asymptotic limit. As
predicted, there is a critical radius beyond which no stationary solutions
exist. The asymptotic predictions shown in Fig.~\ref{fig:bifurcation}
have been translated to the dashed lines of Fig.~\ref{fig.lubri1}. 
These are obtained from (\ref{eq:bifurc}),
where $K_-$ was computed from $h_0$ using (\ref{eq:bessel}). Good
quantitative agreement is achieved for small enough values of the flux.

\begin{figure}
\begin{center}
\includegraphics[width=8.0cm]{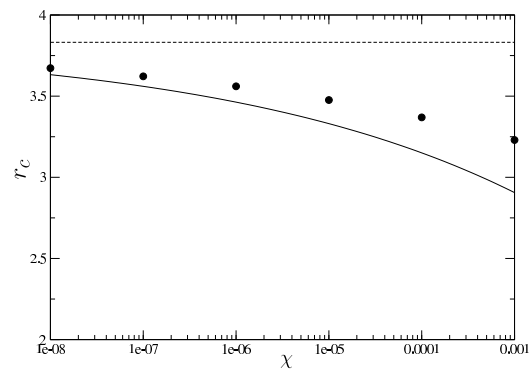}
\caption{Critical radius $r_c$ as a function of the flux $\chi$. The
numerical values obtained from the lubrication equation (dots) indeed
approach the theoretical prediction (\ref{eq.rc}) (solid line). The dashed
line indicates the asymptotic value $r_0=3.8317\cdots$.}.
\label{fig.lubri2}
\end{center}
\end{figure}

Finally, we determined the critical radius $r_c$ for a range of values of
the flux $\chi$. Figure~\ref{fig.lubri2} shows how the numerical values
(dots) indeed approach the asymptotic prediction (solid line) in the limit
of vanishing flux. Due to the very small powers $\chi^{2/15}$, the
convergence towards $r_0=3.8317\cdots$ is extremely slow (horizontal line).
As a consequence,
the correction with respect to this asymptotic value will be significant for
typical experimental values of the flux.

\section{Discussion}

Owing to the smallness of the neck region (\ref{eq.hn}), we can make
the simplification that the pressure inside the gas pocket below the
drop is constant (Fig.~\ref{fig:outer}). This pressure is larger
than the atmospheric pressure and provides the force required to
levitate the drop. Matching the pressure difference across the neck
with the viscous flow then provides the bifurcation diagram of
Fig.~\ref{fig:bifurcation}, yielding a critical neck radius $r_c$.
In the limit of vanishing flux, the critical radius approaches
$r_0=3.8317\cdots$. This value arises because it is the first minimum
of the function characterizing the shape of the gas pocket, which is
the Bessel function $J_0(r)$. For larger $r_n$, the gas pocket shape
would need to become negative, which is of course not allowed.

Experimentally, the size of the drop is measured by looking at the
drop from above. This measurement provides the maximum radius
$r_{\rm max}$ rather than the neck radius, cf. Fig.~\ref{fig:outer}.
For large puddles the difference between $r_{\rm max}$ and $r_n$
approaches $\sqrt{2}-\mathrm{arcccosh} \sqrt{2} \approx 0.53$. For
drop sizes relevant here we confirmed numerically that
$r_{\rm max}-r_n \approx 0.52$. Combined with (\ref{eq.rc}), we thus find

\begin{equation}\label{eq.hieps}
r_{\rm max,st} \approx 4.35 -  1.02 \chi^{1/10}
\end{equation}
for the boundary of stability, expressed in terms of the capillary length.
Typical experimental values of $\chi$ can be extracted using
(\ref{eq.hn}), and typical experiments yield
$h_n\approx 100 \mu\; m$, obtained from diffraction data \cite{Biance03}.
This gives $\chi\approx 10^{-4}$, and thus $r_{\rm max,st} \approx 3.95$,
to be compared to reported experimental values of $4.0\pm 0.2$
\cite{These_Biance,Biance03}. A similar estimate of $\chi$ is obtained
from considering the latent heat of evaporation \cite{Biance03}.
Furthermore, our boundary integral simulations
show that the nonlinear dynamics for larger drops lead to the formation
of chimneys, as observed experimentally. We are therefore confident
that the analysis in terms of Stokes flow provides an accurate
description of this instability.

Let us now return to the argument put forward in
\cite{These_Biance,Biance03}, relating chimney formation to the
Rayleigh-Taylor instability. The later occurs when a layer of fluid
is suspended above another fluid of lower density, so that the system
tends to destabilize due to buoyancy forces. Surface tension opposes
this effect, so that the instability occurs at long wavelengths only.
Biance et al. \cite{These_Biance,Biance03} propose that levitated
drops remain stable as long as axisymmetric perturbations that fit
inside the drop are stable with respect to this buoyancy driven instability.

For an infinitely extended liquid film, one finds that $J_0(kr)$ are
axisymmetric eigenmodes, with the stability criterion $k > 1$. While
the Bessel function does not have a well-defined period, the maximum
drop size was estimated in \cite{These_Biance} by the first minimum
of the mode with $k=1$, occurring at $r_0$. In hindsight, our results
justify this choice of taking the minimum of $J_0(r)$ as the stability
boundary, provided that it is identified with the {\em neck} radius,
rather than with $r_{\rm max}$. With this connection, our results
reduce to the Rayleigh-Taylor argument in the limit of vanishing gas
flow, showing that the balance between buoyancy and surface tension
provides the right mechanism. The effect of the gas flow is to slightly
reduce the range of stable solutions (\ref{eq.hieps}).

\appendix

\section{Gas pocket solution}\label{app.bessel}

In this appendix we expand the gas pocket solution for small
amplitudes and compute the constant $g_2$. We consider the equation

\begin{equation}
\frac{h''}{(1+h'^2)^{3/2}} + \frac{h'}{r(1+h'^2)^{1/2}} + h = c_-,
\end{equation}
with boundary conditions

\begin{eqnarray}
h'(0) &=& 0 \\
h(r_n) &=& 0.
\end{eqnarray}
This is equivalent to solving

\begin{equation}
\frac{y''}{(1+y'^2)^{3/2}} + \frac{y'}{r(1+y'^2)^{1/2}} + y = 0,
\end{equation}
with boundary conditions

\begin{eqnarray}
y'(0) &=& 0 \\
y(r_n) &=& -A,
\end{eqnarray}
where $A=c_-$
We expand in $A$,

\begin{equation}
y(r) = A y_1(r) + A^3 y_3(r) + \mathcal{O}\left( A^5\right).
\end{equation}
This yields a hierarchy of equations

\begin{eqnarray}
y_1'' + \frac{y_1'}{r} + y_1 &=& 0,\\
y_3'' + \frac{y_3'}{r} + y_1 &=& \frac{3}{2} y_1'^2 y_1'' +
\frac{1}{2r} y_1'^3.
\end{eqnarray}
with boundary conditions

\begin{eqnarray}
y_1'(0) &=& 0 \\
y_1(r_n) &=& -1\\
y_3'(0) &=& 0 \\
y_3(r_n) &=& 0.
\end{eqnarray}
The first equation gives $y_1(r)=-J_0(r)/J_0(r_n)$, which can be
inserted into the right hand side of the equation for $y_3(r)$.

To compute the constant $g_2$, we require the ratio $y'(r_0)/y''(r_0)$.
In terms of the expansion

\begin{equation}
\frac{y'(r_0)}{y''(r_0)} = A^2 y_3'(r_0) + \mathcal{O}\left( A^4\right),
\end{equation}
where we used the properties $y_1(r_0)'=0$ and $y_1''(r_0)=-y_1(r_0)=1$.
Comparing to (\ref{eq.gexp}), we simply find

\begin{equation}
g_2 = y_3'(r_0).
\end{equation}
We obtained this value numerically by solving the ODE for $y_3$,
for which we numerically obtained $g_2=1.486\cdots$.

\bibliography{/all_ref}

\begin{thebibliography}{25}
\expandafter\ifx\csname natexlab\endcsname\relax\def\natexlab#1{#1}\fi
\expandafter\ifx\csname bibnamefont\endcsname\relax
  \def\bibnamefont#1{#1}\fi
\expandafter\ifx\csname bibfnamefont\endcsname\relax
  \def\bibfnamefont#1{#1}\fi
\expandafter\ifx\csname citenamefont\endcsname\relax
  \def\citenamefont#1{#1}\fi
\expandafter\ifx\csname url\endcsname\relax
  \def\url#1{\texttt{#1}}\fi
\expandafter\ifx\csname urlprefix\endcsname\relax\def\urlprefix{URL }\fi
\providecommand{\bibinfo}[2]{#2}
\providecommand{\eprint}[2][]{\url{#2}}

\bibitem[{\citenamefont{Duchemin et~al.}(2005)\citenamefont{Duchemin, Lister,
  and Lange}}]{Duchemin05}
\bibinfo{author}{\bibfnamefont{L.}~\bibnamefont{Duchemin}},
  \bibinfo{author}{\bibfnamefont{J.~R.} \bibnamefont{Lister}},
  \bibnamefont{and} \bibinfo{author}{\bibfnamefont{U.}~\bibnamefont{Lange}},
  \bibinfo{journal}{J. Fluid Mech.} \textbf{\bibinfo{volume}{533}},
  \bibinfo{pages}{161} (\bibinfo{year}{2005}).

\bibitem[{\citenamefont{Leidenfrost}(1756)}]{Leidenfrost}
\bibinfo{author}{\bibfnamefont{J.~G.} \bibnamefont{Leidenfrost}},
  \emph{\bibinfo{title}{De aquae communis nonnullis qualitibus tractatus,
  part2}} (\bibinfo{publisher}{Duisburg}, \bibinfo{year}{1756}).

\bibitem[{\citenamefont{Holter and Glasscock}(1952)}]{Holter52}
\bibinfo{author}{\bibfnamefont{N.~J.} \bibnamefont{Holter}} \bibnamefont{and}
  \bibinfo{author}{\bibfnamefont{W.~R.} \bibnamefont{Glasscock}},
  \bibinfo{journal}{J. Acoust. Soc. Am.} \textbf{\bibinfo{volume}{24}},
  \bibinfo{pages}{682} (\bibinfo{year}{1952}).

\bibitem[{\citenamefont{Goldshtick et~al.}(1986)\citenamefont{Goldshtick,
  Khanin, and Ligai}}]{Goldshtik86}
\bibinfo{author}{\bibfnamefont{M.~A.} \bibnamefont{Goldshtick}},
  \bibinfo{author}{\bibfnamefont{V.~M.} \bibnamefont{Khanin}},
  \bibnamefont{and} \bibinfo{author}{\bibfnamefont{V.~G.} \bibnamefont{Ligai}},
  \bibinfo{journal}{J. Fluid Mech.} \textbf{\bibinfo{volume}{166}},
  \bibinfo{pages}{1 } (\bibinfo{year}{1986}).

\bibitem[{\citenamefont{Biance et~al.}(2003)\citenamefont{Biance, Clanet, and
  Qu\'er\'e}}]{Biance03}
\bibinfo{author}{\bibfnamefont{A.~L.} \bibnamefont{Biance}},
  \bibinfo{author}{\bibfnamefont{C.}~\bibnamefont{Clanet}}, \bibnamefont{and}
  \bibinfo{author}{\bibfnamefont{D.}~\bibnamefont{Qu\'er\'e}},
  \bibinfo{journal}{Phys. Fluids} \textbf{\bibinfo{volume}{15}},
  \bibinfo{pages}{1632} (\bibinfo{year}{2003}).

\bibitem[{\citenamefont{Linke et~al.}(2006)\citenamefont{Linke, Alem{\'a}n,
  Melling, Taormina, Francis, Dow-Hygelund, Narayanan, Taylor, and
  Stout}}]{Linke06}
\bibinfo{author}{\bibfnamefont{H.}~\bibnamefont{Linke}},
  \bibinfo{author}{\bibfnamefont{B.~J.} \bibnamefont{Alem{\'a}n}},
  \bibinfo{author}{\bibfnamefont{L.~D.} \bibnamefont{Melling}},
  \bibinfo{author}{\bibfnamefont{M.~J.} \bibnamefont{Taormina}},
  \bibinfo{author}{\bibfnamefont{M.~J.} \bibnamefont{Francis}},
  \bibinfo{author}{\bibfnamefont{C.~C.} \bibnamefont{Dow-Hygelund}},
  \bibinfo{author}{\bibfnamefont{V.}~\bibnamefont{Narayanan}},
  \bibinfo{author}{\bibfnamefont{R.~P.} \bibnamefont{Taylor}},
  \bibnamefont{and} \bibinfo{author}{\bibfnamefont{A.}~\bibnamefont{Stout}},
  \bibinfo{journal}{Phys. Rev. Lett.} \textbf{\bibinfo{volume}{96}},
  \bibinfo{pages}{154502} (\bibinfo{year}{2006}).

\bibitem[{\citenamefont{Couder et~al.}(2005)\citenamefont{Couder, Proti{\`e}re,
  Fort, and Boudaoud}}]{CouderNature}
\bibinfo{author}{\bibfnamefont{Y.}~\bibnamefont{Couder}},
  \bibinfo{author}{\bibfnamefont{S.}~\bibnamefont{Proti{\`e}re}},
  \bibinfo{author}{\bibfnamefont{E.}~\bibnamefont{Fort}}, \bibnamefont{and}
  \bibinfo{author}{\bibfnamefont{A.}~\bibnamefont{Boudaoud}},
  \bibinfo{journal}{Nature} \textbf{\bibinfo{volume}{437}},
  \bibinfo{pages}{208} (\bibinfo{year}{2005}).

\bibitem[{\citenamefont{Gilet et~al.}(2007)\citenamefont{Gilet, Vandewalle, and
  Dorbolo}}]{Gilet07}
\bibinfo{author}{\bibfnamefont{T.}~\bibnamefont{Gilet}},
  \bibinfo{author}{\bibfnamefont{N.}~\bibnamefont{Vandewalle}},
  \bibnamefont{and} \bibinfo{author}{\bibfnamefont{S.}~\bibnamefont{Dorbolo}},
  \bibinfo{journal}{Phys. Rev. E} \textbf{\bibinfo{volume}{76}},
  \bibinfo{pages}{035302(R)} (\bibinfo{year}{2007}).

\bibitem[{\citenamefont{Amarouchene et~al.}(2001)\citenamefont{Amarouchene,
  Cristobal, and Kellay}}]{Yacine01}
\bibinfo{author}{\bibfnamefont{Y.}~\bibnamefont{Amarouchene}},
  \bibinfo{author}{\bibfnamefont{G.}~\bibnamefont{Cristobal}},
  \bibnamefont{and} \bibinfo{author}{\bibfnamefont{H.}~\bibnamefont{Kellay}},
  \bibinfo{journal}{Phys. Rev. Lett.} \textbf{\bibinfo{volume}{95}},
  \bibinfo{pages}{164504} (\bibinfo{year}{2001}).

\bibitem[{\citenamefont{Hervieu et~al.}(2001)\citenamefont{Hervieu, Coutris,
  and Boichon}}]{Hervieu01}
\bibinfo{author}{\bibfnamefont{E.}~\bibnamefont{Hervieu}},
  \bibinfo{author}{\bibfnamefont{N.}~\bibnamefont{Coutris}}, \bibnamefont{and}
  \bibinfo{author}{\bibfnamefont{C.}~\bibnamefont{Boichon}},
  \bibinfo{journal}{Nucl. Eng. Design} \textbf{\bibinfo{volume}{204}},
  \bibinfo{pages}{167} (\bibinfo{year}{2001}).

\bibitem[{\citenamefont{Biance}(2003)}]{These_Biance}
\bibinfo{author}{\bibfnamefont{A.}~\bibnamefont{Biance}},
  \emph{\bibinfo{title}{Gouttes inertielles: de la cal\'efaction a
  l'\'etalement}} (\bibinfo{year}{2003}).

\bibitem[{\citenamefont{Watchers et~al.}(1966)\citenamefont{Watchers, Bonne,
  and van Nouhuis}}]{Watchers66}
\bibinfo{author}{\bibfnamefont{L.~H.~J.} \bibnamefont{Watchers}},
  \bibinfo{author}{\bibfnamefont{H.}~\bibnamefont{Bonne}}, \bibnamefont{and}
  \bibinfo{author}{\bibfnamefont{H.~J.} \bibnamefont{van Nouhuis}},
  \bibinfo{journal}{Chem. Eng. Sci.} \textbf{\bibinfo{volume}{21}},
  \bibinfo{pages}{923} (\bibinfo{year}{1966}).

\bibitem[{\citenamefont{Adachi and Takaki}(1984)}]{Takaki84}
\bibinfo{author}{\bibfnamefont{K.}~\bibnamefont{Adachi}} \bibnamefont{and}
  \bibinfo{author}{\bibfnamefont{R.}~\bibnamefont{Takaki}},
  \bibinfo{journal}{J. Phys. Soc. Jap.} \textbf{\bibinfo{volume}{53}},
  \bibinfo{pages}{4184} (\bibinfo{year}{1984}).

\bibitem[{\citenamefont{Takaki and Adachi}(1985)}]{Takaki85}
\bibinfo{author}{\bibfnamefont{R.}~\bibnamefont{Takaki}} \bibnamefont{and}
  \bibinfo{author}{\bibfnamefont{K.}~\bibnamefont{Adachi}},
  \bibinfo{journal}{J. Phys. Soc. Jap.} \textbf{\bibinfo{volume}{54}},
  \bibinfo{pages}{2462} (\bibinfo{year}{1985}).

\bibitem[{\citenamefont{Tokugawa and Takaki}(1994)}]{Tokugawa94}
\bibinfo{author}{\bibfnamefont{N.}~\bibnamefont{Tokugawa}} \bibnamefont{and}
  \bibinfo{author}{\bibfnamefont{R.}~\bibnamefont{Takaki}},
  \bibinfo{journal}{J. Phys. Soc. Jap.} \textbf{\bibinfo{volume}{63}},
  \bibinfo{pages}{1758} (\bibinfo{year}{1994}).

\bibitem[{\citenamefont{Strier et~al.}(2000)\citenamefont{Strier, Duarte,
  Ferrari, and Mindlin}}]{Strier00}
\bibinfo{author}{\bibfnamefont{D.~E.} \bibnamefont{Strier}},
  \bibinfo{author}{\bibfnamefont{A.~A.} \bibnamefont{Duarte}},
  \bibinfo{author}{\bibfnamefont{H.}~\bibnamefont{Ferrari}}, \bibnamefont{and}
  \bibinfo{author}{\bibfnamefont{G.}~\bibnamefont{Mindlin}},
  \bibinfo{journal}{Physica A} \textbf{\bibinfo{volume}{283}},
  \bibinfo{pages}{261} (\bibinfo{year}{2000}).

\bibitem[{\citenamefont{Snezhko et~al.}(2008)\citenamefont{Snezhko, {Ben
  Jacob}, and Aranson}}]{Aranson08}
\bibinfo{author}{\bibfnamefont{A.}~\bibnamefont{Snezhko}},
  \bibinfo{author}{\bibfnamefont{E.}~\bibnamefont{{Ben Jacob}}},
  \bibnamefont{and} \bibinfo{author}{\bibfnamefont{I.~S.}
  \bibnamefont{Aranson}}, \bibinfo{journal}{New J. Phys.}
  \textbf{\bibinfo{volume}{10}}, \bibinfo{pages}{043034}
  (\bibinfo{year}{2008}).

\bibitem[{\citenamefont{Yoshiyasu et~al.}(1996)\citenamefont{Yoshiyasu,
  Matsuda, and Takaki}}]{Yoshiyasu96}
\bibinfo{author}{\bibfnamefont{N.}~\bibnamefont{Yoshiyasu}},
  \bibinfo{author}{\bibfnamefont{K.}~\bibnamefont{Matsuda}}, \bibnamefont{and}
  \bibinfo{author}{\bibfnamefont{R.}~\bibnamefont{Takaki}},
  \bibinfo{journal}{J. Phys. Soc. Jap.} \textbf{\bibinfo{volume}{65}},
  \bibinfo{pages}{2068} (\bibinfo{year}{1996}).

\bibitem[{\citenamefont{Batchelor}(1967)}]{Batch67}
\bibinfo{author}{\bibfnamefont{G.~K.} \bibnamefont{Batchelor}},
  \emph{\bibinfo{title}{An introduction to Fluid Dynamics}}
  (\bibinfo{publisher}{Cambridge University Press}, \bibinfo{year}{1967}).

\bibitem[{\citenamefont{Rallison and Acrivos}(1978)}]{RA78}
\bibinfo{author}{\bibfnamefont{J.~M.} \bibnamefont{Rallison}} \bibnamefont{and}
  \bibinfo{author}{\bibfnamefont{A.}~\bibnamefont{Acrivos}},
  \bibinfo{journal}{J. Fluid Mech.} \textbf{\bibinfo{volume}{89}},
  \bibinfo{pages}{191} (\bibinfo{year}{1978}).

\bibitem[{\citenamefont{Happel and Brenner}(1983)}]{HB83}
\bibinfo{author}{\bibfnamefont{J.}~\bibnamefont{Happel}} \bibnamefont{and}
  \bibinfo{author}{\bibfnamefont{H.}~\bibnamefont{Brenner}},
  \emph{\bibinfo{title}{Low Reynolds Number Hydrodynamics}}
  (\bibinfo{publisher}{Martinus Nijhoff Publishers}, \bibinfo{year}{1983}).

\bibitem[{\citenamefont{Blake}(1972)}]{Blake71}
\bibinfo{author}{\bibfnamefont{J.~R.} \bibnamefont{Blake}},
  \bibinfo{journal}{Proc. Camb. Phil. Soc} \textbf{\bibinfo{volume}{70}},
  \bibinfo{pages}{303} (\bibinfo{year}{1972}).

\bibitem[{\citenamefont{Pozrikidis}(1990)}]{P90}
\bibinfo{author}{\bibfnamefont{C.}~\bibnamefont{Pozrikidis}},
  \bibinfo{journal}{J. Fluid Mech.} \textbf{\bibinfo{volume}{215}},
  \bibinfo{pages}{331} (\bibinfo{year}{1990}).

\bibitem[{\citenamefont{Eggers et~al.}(1999)\citenamefont{Eggers, Lister, and
  Stone}}]{ELS99}
\bibinfo{author}{\bibfnamefont{J.}~\bibnamefont{Eggers}},
  \bibinfo{author}{\bibfnamefont{J.~R.} \bibnamefont{Lister}},
  \bibnamefont{and} \bibinfo{author}{\bibfnamefont{H.~A.} \bibnamefont{Stone}},
  \bibinfo{journal}{J. Fluid Mech.} \textbf{\bibinfo{volume}{401}},
  \bibinfo{pages}{293} (\bibinfo{year}{1999}).

\bibitem[{\citenamefont{Bart}(1968)}]{B68}
\bibinfo{author}{\bibfnamefont{E.}~\bibnamefont{Bart}}, \bibinfo{journal}{Chem.
  Engin. Sci.} \textbf{\bibinfo{volume}{23}}, \bibinfo{pages}{193}
  (\bibinfo{year}{1968}).

\end{thebibliography}
\end{document}